\documentclass[10pt,journal,compsoc]{IEEEtran}

\pdfoutput=1

\ifCLASSOPTIONcompsoc
  
  \usepackage[nocompress]{cite}
\else
  \usepackage{cite}
\fi

\usepackage{amsmath}
\usepackage{csquotes}
\usepackage{multirow}
\usepackage{pbox}
\usepackage{braket}
\usepackage{graphics}
\usepackage{graphicx}
\usepackage{url}
\usepackage{comment}
\usepackage{booktabs}
\usepackage[ruled,linesnumbered]{algorithm2e}
\usepackage{balance}
\usepackage{xcolor}
\usepackage{multirow}\usepackage[numbers,sort]{natbib}
\usepackage{graphicx}
\usepackage{textcomp}
\usepackage{newunicodechar}
\usepackage{forest}
\usepackage{multirow}
\usepackage{fixmath}
\usepackage{etoolbox}
\usepackage{multicol} 
\usepackage{authblk}
\usepackage{comment} 
\usepackage{graphicx}
\usepackage{tabularray}
\usepackage{subcaption}
\usepackage{ulem}
\usepackage{array}
\usepackage{multirow}
\usepackage{ragged2e}
\usepackage{amsmath}
\usepackage{academicons}
\usepackage[utf8]{inputenc}
\usepackage[misc]{ifsym}
\usepackage{etoolbox}

\usepackage{pifont}
\usepackage{csquotes}
\newcommand{\xmark}{\ding{55}}
\newcommand{\cmark}{\ding{51}}
\usepackage{tcolorbox}
\usepackage{xspace}
\usepackage{seqsplit}
\usepackage{diagbox}
\ifCLASSINFOpdf
\else
\fi

\begin{document}

\title{IoTFuzzSentry: A Protocol Guided Mutation Based Fuzzer for Automatic Vulnerability Testing in Commercial IoT Devices}

\author{Priyanka Rushikesh Chaudhary 
        and Rajib Ranjan Maiti}

\IEEEtitleabstractindextext{%
\begin{abstract}
  Protocol fuzzing is a scalable and cost-effective technique for identifying security vulnerabilities in deployed Internet of Things (IoT) devices. During their operational phase, IoT devices often run lightweight servers to handle user interactions, such as video streaming or image capture in smart cameras. Implementation flaws in transport or application-layer security mechanisms can expose IoT devices to a range of threats, including unauthorized access and data leakage. This paper addresses the challenge of uncovering such vulnerabilities by leveraging protocol fuzzing techniques that inject crafted transport and application-layer packets into IoT communications. 
We present a mutation-based fuzzing tool, named \textit{IoTFuzzSentry}, to identify specific non-trivial vulnerabilities in commercial IoT devices. We further demonstrate how these vulnerabilities can be exploited in real-world scenarios. We integrated our fuzzing tool into a well-known testing tool \textit{Cotopaxi} and evaluated it with commercial-off-the-shelf (COTS) IoT devices such as IP cameras and Smart Plug. 
Our evaluation revealed vulnerabilities categorized into four types (IoT Access Credential Leakage, Sneak IoT Live Video Stream, Creep IoT Live Image, IoT Command Injection) and we show their exploits using three IoT devices. We have responsibly disclosed all these vulnerabilities to the respective vendors. 
So far, we have published two CVEs, CVE-2024-41623 and CVE-2024-42531, and one is awaiting. To extend the applicability of IoTFuzzSentry, we have investigated the traffic of six additional IoT devices and our analysis shows that these devices can have similar vulnerabilities, due to the presence of a similar set of application protocols.
We believe that our IoTFuzzSentry has the potential to discover unconventional security threats and allow IoT vendors to strengthen the security of their commercialized IoT devices automatically with negligible overhead.

\end{abstract}

}

\maketitle

\IEEEdisplaynontitleabstractindextext

\IEEEpeerreviewmaketitle

\IEEEraisesectionheading{\section{Introduction}\label{sec:introduction}}

The deployment of commercial Internet of Things (IoT) devices is growing rapidly, and this trend is expected to continue until 2028 \cite{IoTMarket1}. Due to intense market competition, the IoT vendors primarily focus on the time to market and business outcome \cite{hassan2019Elsevier, johnson2020PLOSArticle, fichtner2022Journal}. Hence, IoT developers work primarily on the correctness of device functionalities, ignoring one of the critical aspects of cybersecurity. Post-deployment, it is often difficult to update or install any patch into the device \cite{Neshenko2019DemystifyingISIEEE, haddadpajouh2021ElsevierIoT}. Hence, such devices become lucrative targets for cyber hackers, leading to wealth loss and defamation \cite{kalbo2020securityMDPISensors, VerkadaAttack, kim2011Elsevier}. Therefore, it is of paramount importance to uncover the device vulnerabilities at the earliest. This paper presents an approach for automatically discovering vulnerabilities in IoT devices using protocol-guided lexical fuzzing.

Once setup is done, an IoT device starts communicating with its associated cloud server and a mobile application, hereafter referred to as the \textit{controller}, installed on the user's smartphone. Typically, an IoT device runs a lightweight server that waits for user commands via the controller; once received, it executes the command, possibly sends back a reply packet to the controller, and communicates with its associated server. 
For example, an IoT camera sends its video stream to its controller after setup, via a cloud server. Further, the user can send commands to move its focus area or take a snap from its live video, referred to as a \textit{live image}. 
An IoT device may use multiple application protocols to offer such services. For example, video streaming can be delivered over Real Time Streaming Protocol (RTSP), and the user commands may be transferred using simple Hypertext Transfer Protocol (HTTP). The protocols used by IoT devices often have no or limited security features. For example, HTTP GET requests can have an optional header field of \textit{authorization} to authenticate their payload. However, due to the differences or inherent flaws in implementing such provisions, commercial IoT devices may suffer severe vulnerabilities \cite{zhou2019USENIX}. 
We aim to effectively address the problem of automatically discovering such vulnerabilities in operational IoT devices by using protocol-guided mutation-based lexical fuzzing.

Discovering the vulnerabilities in IoT devices has gained significant momentum in recent times.
While attackers are interested in exploiting the vulnerabilities in commercial IoT devices, possibly for economic gains \cite{VerkadaAttack, ZewenICST2024},
security researchers are interested in discovering them early, e.g., by using fuzzing \cite{li2018fuzzingSpringer, mallissery2023demystifyFuzzingSurvey, manes2018fuzzingSurvey}.
Fuzzing can discover security threats in firmware \cite{AotFirmware2023Eurosnp, zhang2021SIoTfuzzerMDPI, feng2021snipuzzACM}, language-specific libraries \cite{Libfuzzer}, 
and IoT devices \cite{redini2021dianeIEEESnP}.  
In particular, prior works \cite{xu2022fiotfuzzerIEEEACIS, chen2018iotfuzzerNDSS} demonstrate that fuzzing can discover security flaws in the controller or the cloud server of an IoT device when appropriate seed input is in place. We broadly position our work in the category of bug hunting using fuzzing. 
However, in contrast to prior works on IoT fuzzing, we aim to automatically 
discover non-crash vulnerabilities, thus potentially uncovering attacks beyond denial-of-service. 
Moreover, our current focus is to discover these vulnerabilities based on a single protocol-specific packet rather than using an appropriate sequence of packets.

First, we collected network traces of several commercially available IoT devices in our laboratory setup and inspected the data being exchanged using the protocols above the transport layer. 
Because the aim is to discover non-crash vulnerabilities, we have consulted the NIST draft "Baseline Security Criteria for Consumer IoT devices" \cite{NISTDraft2021SecurityIoT} to look for confirmation of the baseline IoT security guidelines.
Based on the guidelines under the columns of "Product configuration", "Data Protection" and "Logical Access To Interfaces", we pose a few simple security research questions, \textit{Is it possible to discover IoT device credentials?} \textit{Is there any pair of request/response packets that triggers the release of device credential?} \textit{Can an IoT device accept a user command when placed in a fuzzed packet?} \textit{Can existing fuzzing attacks be applied to IoT devices of similar types?} 
One needs to overcome a few challenges to answer these questions.
First, discover only those packets among a large number of network packets that carry a specific type of security parameter, such as plaintext or encoded or encrypted user credentials, user commands, and message authentication code, if any.
Second, if the payload is encoded or encrypted, then discover the exact scheme by applying some heuristic or social engineering so that an arbitrary command can be encoded or encrypted. 
Third, identify the exact fields in the payload that can be subject to mutation-based lexical fuzzing. 
Overcoming these challenges can certainly lead us to discover any potential vulnerability related to data security and access, like IoT credentials, video or image files, and device control, rather than conventional device crashes.

To address the challenges mentioned in the preceding paragraph, we extend our basic idea proposed in \cite{Priyanka2024CODASPY}. We design and implement a customized mutation-based lexical fuzzer 
\textit{IoTFuzzSentry} to automatically identify implementation-specific IoT vulnerabilities beyond crashes.
\textit{IoTFuzzSentry} extracts a limited set of seed inputs from the legitimate network traffic of IoT devices under test (IUT). These seed inputs are then used to create a larger set of mutated inputs, which are then placed in certain fields of crafted packets of a particular application protocol. These packets are then injected into the IUT to discover specific types of vulnerabilities. Finally, the \textit{IoTFuzzSentry} generates a descriptive report based on the responses received, reflecting the test statistics and performance.  

To enhance the usability and applicability of our approach, we have integrated \textit{IoTFuzzSentry} 
with a well-known mutation-based IoT fuzzer called Cotopaxi \cite{cotopaxi}. Currently, Cotopaxi supports 14 IoT application protocols for fuzzing several IoT devices. 
However, existing Cotopaxi payloads are effective only for specific IoT cameras (Foscam, Beward, TP-Link) and rely on known URIs. Our work extends Cotopaxi’s capability.

We have applied \textit{IoTFuzzSentry} on three commercially available IoT devices: the D3D IoT camera, Ezviz IoT camera, and TP-Link Smart plug. We have discovered three major vulnerabilities in these devices, for which we have received two published CVE IDs, and one is being patched. 
Our contributions are as follows:
\begin{enumerate}
    \item We design and implement \textit{IoTFuzzSentry}\footnote{We make the code base of IoTFuzzSentry, along with sample pcap files, available online \cite{githubRepo}} a protocol-guided, mutation-based lexical fuzzer that discovers vulnerabilities beyond crashes in commercial IoT devices.
    
    \item  We evaluate \textit{IoTFuzzSentry} on three COTS IoT devices, i.e., D3D IoT camera, Ezviz IoT camera, and TP-Link Smart plug.
    Our evaluation revealed seven vulnerabilities that, instead of crashing the devices, can cause severe privacy and cyber threats. 
       
    \item We show that the existing HTTP and RTSP-based CVEs related to IoT cameras do not apply to the devices considered in this paper. Hence, we have published two new CVEs (CVE-2024-41623 for D3D and CVE-2024-42531 for Ezviz). 
    The plug's vulnerability is related to TP-Link SmartHome/JSON's proprietary protocol, and the vendor has requested some time to patch it before reporting it through CVE. 

    \item Using six additional IoT devices from different vendors, we have discovered that these devices also use a similar set of application protocols, i.e., HTTP/TCP/RTSP, and the payload exhibits a certain directory structure indicating the exact files being transferred in plaintext. Thus, though not included, IoTFuzzSentry can be easily extended to test vulnerabilities like credential leakage, creeping live image, command injection, and sneaking live video in these IoT devices.
    
    \item To extend applicability, \textit{IoTFuzzSentry} is integrated with a well-known mutation-based fuzzer called Cotopaxi after passing the necessary quality checks. Thus, we have enhanced the capability of Cotopaxi by including three new IoT devices and seven new vulnerabilities. Our analysis shows that the new integration performs as per the existing Cotopaxi.
\end{enumerate}

The rest of the paper is organized as follows. Section \ref{sec:relwork} describes related works, and Section \ref{sec:systhrtmodel} presents the system and threat model. Section \ref{sec:IoTFuzzSentry} details the design and implementation of \textit{IoTFuzzSentry}. Section \ref{sec:integratoin-into-cotopaxi} describes the details of Cotopaxi and the new integration of \textit{IoTFuzzSentry}. Section \ref{sec:effect_fuzzer} shows the effectiveness of our \textit{IoTFuzzSentry}, and Section \ref{sec:conclusion} concludes the work in this paper.
    
\section{Related work} 
\label{sec:relwork}

\subsection{System Testing and Fuzz Testing}
Traditionally, the methods for discovering a vulnerability in a system being designed or implemented include static analysis \cite{sachidananda2020ACM, ferrara2021Springer}, dynamic analysis \cite{ozmen2022ACMSIGSAC, Yu12020MDPIFutureInternet}, taint analysis \cite{yavuz2022CODASPYy, mandal2020ACMSymposium}, and symbolic execution \cite{stephens2016NDSS, ognawala2017Arxiv, Ognawala2018ACM}. One of the powerful methods for discovering vulnerability in an operational system is fuzzing \cite{Eceiza2021IEEEIoTJournal, li2018fuzzingSpringer} due to a higher code coverage \cite{li2018fuzzingSpringer}, \cite{Luo2021ACSAC, Chenyang2023USENIX, xumock2024NDSS}.
In its basic form, a fuzzer works by providing some unexpected inputs into a system under test and monitoring its response; the more anomalies or unexpected behaviors in the response, the higher the effectiveness of fuzzing. 
Based on the kind of knowledge used in generating the inputs, fuzzing can be classified into two types: generation-based and mutation-based \cite{li2018fuzzingSpringer}. In the generation-based fuzzing \cite{Situ2023IoTJ}, \cite{Blair2022ACMTPS}, the inputs are produced based on the knowledge about the target being tested, e.g., a high-level specification of the system or a release of an abstract code base. In mutation-based fuzzing, the inputs are produced by altering a subset of bits in known inputs, and therefore, it requires no or limited knowledge of the target. 
Since its inception, fuzzing has been used to identify vulnerabilities in large-scale programs. Various fuzzing techniques are employed in practice based on the features of target applications. 
For example, fuzzing an application that works on specific types of files can be done by \textit{File Format fuzzing} using tools like \textit{Peach} and \textit{AFL} and its extensions. Researchers use \textit{KAFL} \cite{schumilo2017kaflUSENIX}, \cite{wang2023arxiv} and \textit{Syzkaller} \cite{Syzkaller} to fuzz operating system kernels. A software that implements a communication protocol can be fuzzed using \textit{SPIKE} \cite{SPIKE}, \textit{AutoFuzz} \cite{gorbunov2010autofuzzIJCSNS} and \textit{SNOOZE} \cite{SNOOZE2006Springer}. Overall, fuzzing has been effective in discovering vulnerabilities in a wide variety of traditional systems \cite{Sorniotti2023IEEE}, \cite{Shashank2024ASIACCS}, and we focus on the fuzzing of IoT systems.

\begin{table*}[htbp]
\centering
\caption{Comparisons of Mutation-based Fuzzers for IoT Devices}
\label{tab:compfuzz}
\begin{tabular}{|c|c|c|cccc|}
\hline
\multirow{2}{*}{\begin{tabular}[c]{@{}c@{}}Mutation-based \\ Fuzzers\end{tabular}} &
  \multirow{2}{*}{\begin{tabular}[c]{@{}c@{}}Fuzz \\ Layer\end{tabular}} &
  \multirow{2}{*}{\begin{tabular}[c]{@{}c@{}}Fuzzing\\ Applications\end{tabular}} &
  \multicolumn{4}{c|}{Mutations Using Message Fragments} \\ \cline{4-7} 
 &
   &
   &
  \multicolumn{1}{c|}{\begin{tabular}[c]{@{}c@{}}URLs to control\\ IP camera\end{tabular}} &
  \multicolumn{1}{c|}{\begin{tabular}[c]{@{}c@{}}Live Image\\ with path\end{tabular}} &
  \multicolumn{1}{c|}{\begin{tabular}[c]{@{}c@{}}Directory paths \\of media files\end{tabular}} &
  \begin{tabular}[c]{@{}l@{}}Control\\ Commands\end{tabular} \\ \hline
AFL \cite{AFL} &
  ---- &
  Softwares &
  \multicolumn{1}{c|}{\xmark} &
  \multicolumn{1}{c|}{\xmark} &
  \multicolumn{1}{c|}{\xmark} & \xmark
   \\ \hline
LibFuzzer \cite{Libfuzzer}&
  ---- &
  \begin{tabular}[c]{@{}c@{}}Libraries,\\ Dictionary\end{tabular} &
  \multicolumn{1}{c|}{\xmark} &
  \multicolumn{1}{c|}{\xmark} &
  \multicolumn{1}{c|}{\xmark} &  \xmark
   \\ \hline
RPFuzzer \cite{wang2013rpfuzzer} &
  \begin{tabular}[c]{@{}c@{}}Routing \\ Protocol\end{tabular} &
  Routers &
  \multicolumn{1}{c|}{\xmark} &
  \multicolumn{1}{c|}{\xmark} &
  \multicolumn{1}{c|}{\xmark} &  \xmark
   \\ \hline
FIoTFuzzer \cite{xu2022fiotfuzzerIEEEACIS} &
  \begin{tabular}[c]{@{}c@{}}Network and \\ above layer\end{tabular} &
  \begin{tabular}[c]{@{}c@{}}IP cameras,\\ Routers, Smart Bulb\end{tabular} &
  \multicolumn{1}{c|}{\xmark} &
  \multicolumn{1}{c|}{\xmark} &
  \multicolumn{1}{c|}{\xmark} & \xmark
   \\ \hline
IoTFuzzer \cite{chen2018iotfuzzerNDSS} &
  Network layer &
  Mobile App &
  \multicolumn{1}{c|}{\xmark} &
  \multicolumn{1}{c|}{\xmark} &
  \multicolumn{1}{c|}{\xmark} &  \xmark
   \\ \hline
DIANE \cite{redini2021dianeIEEESnP} &
  \begin{tabular}[c]{@{}c@{}}Network and \\ above layer\end{tabular} &
  Mobile App &
  \multicolumn{1}{c|}{\xmark} &
  \multicolumn{1}{c|}{\xmark} &
  \multicolumn{1}{c|}{\xmark} &  \xmark
   \\ \hline
SIoTFuzzer \cite{zhang2021SIoTfuzzerMDPI} &
  \begin{tabular}[c]{@{}c@{}}Network\\ Layer\end{tabular} &
  Web Interface &
  \multicolumn{1}{c|}{\xmark} &
  \multicolumn{1}{c|}{\xmark} &
  \multicolumn{1}{c|}{\xmark} & \xmark
   \\ \hline
Cotopaxi \cite{cotopaxi} &
  \begin{tabular}[c]{@{}c@{}}Network and \\ above layer\end{tabular} & \begin{tabular}[c]{@{}c@{}}
  IoT devices, \\ Protocol Servers \end{tabular} &
  \multicolumn{1}{c|}{\xmark} &
  \multicolumn{1}{c|}{\xmark} &
  \multicolumn{1}{c|}{\xmark} & \xmark
   \\ \hline   
Our Approach &
  \begin{tabular}[c]{@{}c@{}}Network and \\ above layer\end{tabular} &
  \begin{tabular}[c]{@{}c@{}}IP cameras, \\ Smart Plugs,\end{tabular} &
  \multicolumn{1}{c|}{\cmark} &
  \multicolumn{1}{c|}{\cmark} &
  \multicolumn{1}{c|}{\cmark} & \cmark
   \\ \hline
\end{tabular}
\end{table*}

\subsection{Generation-based IoT Fuzzer}
Generation-based fuzzing tools, like Sulley \cite{Sulley} and Boofuzz \cite{boofuzz}, automatically generate numerous test cases, reducing manual intervention, and are best suited for large systems with structured input formats. These tools continuously monitor the system under test using behavioral aspects. Basically, such tools focus on reducing the manual effort required to carry out the tests. This fuzzing is generally applied to large software systems that handle highly structured input \cite{Kyriakos2020USENIX}. 
Due to limited visibility into device internals during runtime, we consider mutation-based lexical fuzzers to be more relevant to our work in this paper. 

\subsection{Mutation-based IoT Fuzzer}
In general, a mutation-based fuzzer uses a small set of valid inputs as a seed and then creates a large set of inputs by mutating them. 
In practice, often a generation-based fuzzer integrates additional modules to also perform mutation-based fuzzing. For example, a version of \textit{AFL} supports mutation-based fuzzing. Other tools, like LibFuzzer \cite{Libfuzzer}, primarily support mutation-based fuzzing. 
\textit{Snipuzz} \cite{feng2021snipuzzACM} starts by creating a number of initial test cases by considering the system under test (SUT) as a black box and then creates a large set of inputs by considering the response generated by the initial inputs, thus creating a feedback loop for test case generation. 

However, this fuzzing technique creates a large number of inputs that may not be accepted by the SUT, e.g., if the SUT requires a particular grammar or structure in the input, like a protocol packet format. 
\textit{FIoTFuzzer} in \cite{xu2022fiotfuzzerIEEEACIS} requires mutating message fragments that an IoT device may accept. 
\textit{IoTFuzzer} \cite{chen2018iotfuzzerNDSS} detects memory faults in the IoT device via its controller, where it mutates specific fields in the communication packets. 
\textit{DIANE} \cite{redini2021dianeIEEESnP} identifies code segments in the controller, where each input is first validated based on the required format, and then code coverage is measured. 
Further, the techniques in \cite{wang2019discoveringHindawi} have proposed to build a weighted message parse tree (WMPT) to direct the process of creating mutated inputs based on a pre-defined structure, and then the inputs are applied to fuzz UI components in the web interface of an IoT device. 
\textit{SIoTFuzzer} \cite{zhang2021SIoTfuzzerMDPI} is yet another tool for fuzzing the web interfaces of a number of IoT devices, where a special function, called \textit{stateful message generation} (SMG), helps in deciding the likelihood that devices may receive the mutated messages. 
Among the most closely related works, \textit{Cotopaxi} \cite{cotopaxi} is a Python-based open-sourced simple tool that is closer to our work, and we have considered having our proposed IoTFuzzSentry integrated into it. 
Table \ref{tab:compfuzz} summarizes recent mutation-based fuzzers relevant to IoT security testing, including our proposed IoTFuzzSentry.

\section{System and Threat Model}
\label{sec:systhrtmodel}
\begin{figure}[htbp]
   \centering
   \includegraphics[width=\linewidth]{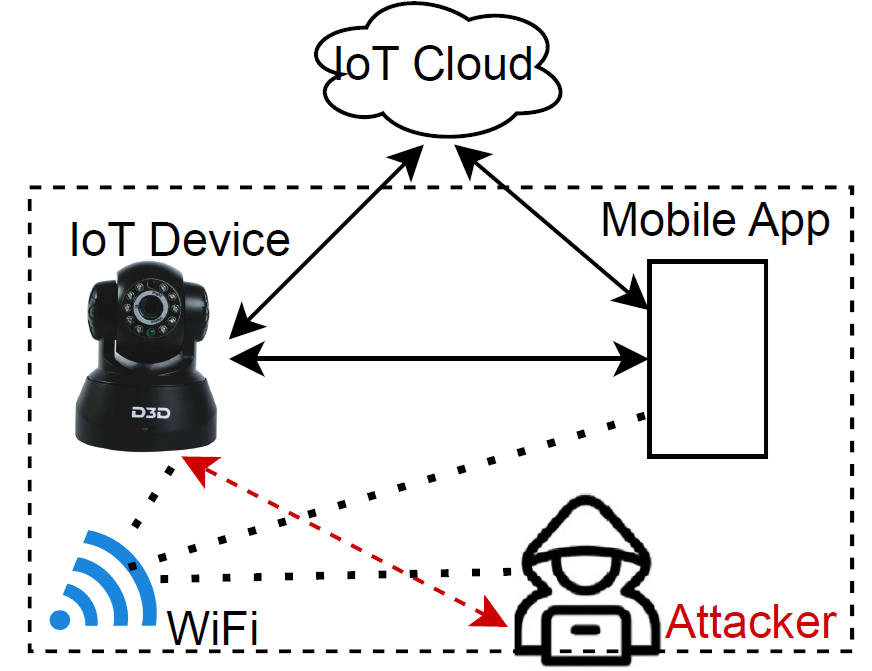}
   \caption{System Model of IoTFuzzSentry} 
    \label{fig:sysmodel}
\end{figure} 

Our proposed IoTFuzzSentry is built assuming a security testing environment in a smart home-like scenario, where the IoT devices are operational and can access the Internet via a Wi-Fi router. Every IoT device has its controller, which does not need to be connected to the router to access the IoT devices.  Figure \ref{fig:sysmodel} shows a simplified system model of  \textit{IoTFuzzSentry}. 
IoTFuzzSentry is used by the security testing team vis-a-vis an attacker. The controller is either connected to the Wi-Fi router at the smart home, or it is remotely operating the IoT device.  
The controller may not detect the presence of the attacker that interacts with the IoT device using IoTFuzzSentry. 


We assume that an attacker is an active attacker who can proactively inject network packets so that the IoT device under test (IUT) can accept them. Essentially, the attacker has access to the network, wired or wireless, that the IoT device is connected to, and the attacker can passively sniff network communication packets of the IUT. However, the attacker neither tries to gain access to the companion mobile application, i.e., the controller, nor requires any access to it. Also, the attacker does not interact with the associated cloud server of IUT, nor does it explore any vulnerability in this server. 
To force the IoT device to accept the fuzzed network packets, the attacker may impersonate the controller by assuming its IP address, if required. Further, the attacker has compatible hardware, like Wi-Fi or Ethernet adapters, to execute IoTFuzzSentry. In other words, the attacker is capable of injecting network packets into the network of the targeted IoT device.

\section{IoTFuzzSentry: The Fuzzing Tool}
\label{sec:IoTFuzzSentry}

\begin{figure}[h]
   \centering
   \includegraphics[width=\linewidth]{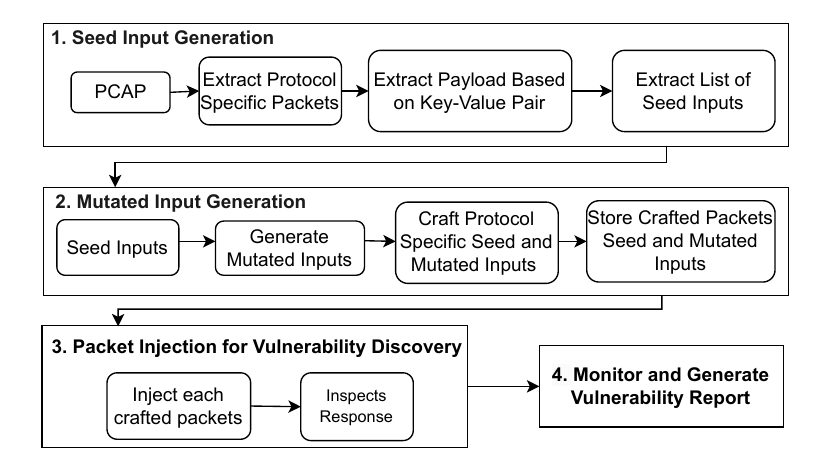}
   \caption{Overview of IoTFuzzSentry.}
    \label{fig:designmodel}
\end{figure}

A detailed design of \textit{IoTFuzzSentry} is shown in Figure \ref{fig:designmodel}. 
It includes four modules, i) seed input collector: uses network traffic of IUT collected passively, ii) mutated input generator: uses seed inputs, iii) fuzzed packet injector: crafts packets using mutated payloads and sends to IUT, and iv) vulnerability assessor: monitors responses and prepares vulnerability report. 
We note that, unlike several other fuzzing techniques, our tool does not 
involve any instrumentation, and it can be launched out-of-the-box. 
In the subsequent sections, we discuss the design details for each \textit{IoTFuzzSentry} module.

\subsection{Seed Input and Mutation}
Commercially available IoT devices can use a variety of application protocols, from standard protocols like HTTP and RTSP to proprietary protocols like TP-Link SmartHome/JSON. In general, these protocols do not use encryption, and hence, both the header and payload can be easily intercepted. 
Device-specific application packets that carry either commands or IoT data are considered as input to our fuzzer. 
Our fuzzer broadly falls under the category of lexical fuzzes, where the field-specific values are fuzzed while the structural properties remain unchanged. 
Note that lexical fuzzing can have three other variations: random fuzzing, grey-box fuzzing, and search-based fuzzing. 
This type of mutation-based fuzzing is fundamentally different from several existing fuzzers, like AFLNet, that rely on altering a known sequence of packets that aim to discover crash. 
Our aim is to discover certain vulnerabilities that, instead of crashing, either exfiltrate sensitive data or allow an attacker to access it. 
\begin{figure}[hbtp]
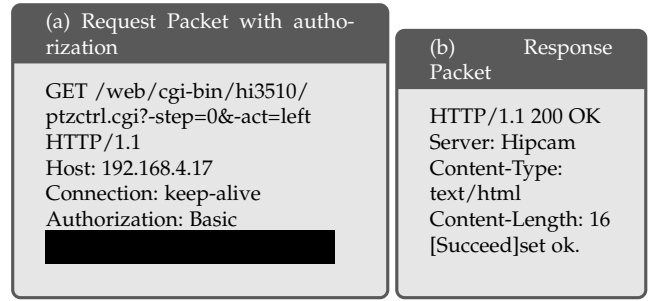

\centering
\begin{frame}

\resizebox{\linewidth}{!}{
\begin{tabular}{ccc}
\begin{tcolorbox}
[colback=gray!20!white,colframe=gray!70!black,width=6cm, title= (a) Request Packet with authorization]
GET  /web/cgi-bin/hi3510/\\ptzctrl.cgi?-step=0\&-act=left HTTP/1.1\\
Host: 192.168.4.17\\
Connection: keep-alive\\
Authorization: Basic\\\colorbox{black}{YWRtaW46YWRtaW4xMjM=}\\
\end{tcolorbox}
\begin{tcolorbox}[colback=gray!20!white,colframe=gray!70!black, width=4cm,  title=(b) Response Packet]
HTTP/1.1 200 OK\\
Server:  Hipcam\\
Content-Type: text/html\\
Content-Length:  16\\
\lbrack Succeed\rbrack  set ok.\\
\end{tcolorbox}
\end{tabular}}
\end{frame}

\caption{HTTP Packets with authorization header in D3D Camera.}
   \label{fig:d3dHttpPkt}
\end{figure}
\subsubsection{Fuzzing HTTP Packets} 
Figure \ref{fig:d3dHttpPkt} shows an example of a HTTP request packet sent to D3D IoT camera and the corresponding response packet from it. 
The request packet actually contains a command to turn the camera's focus to its left.
The shaded text is the value of the \textit{authorization} tag. This tag indicates that the command within \textit{GET} method is authorized. 
To check any known vulnerability, we perform social engineering about this tag in this camera and we found that the value of this tag is a \textit{base64} string, decoding which discloses the username and password being carried in the packet (this is in line with the revelation in \cite{rfc2617HTTPAuth}). 
Because a particular username and password are set during the provisioning phase of the device, ideally, the value of this tag in the operational phase is not subject to mutation. In other words, fuzzing this tag might result in a report stating the device is not vulnerable, and hence, we avoid fuzzing the value in this field.

\begin{figure}[h]
   \centering
   \includegraphics[width=\linewidth]{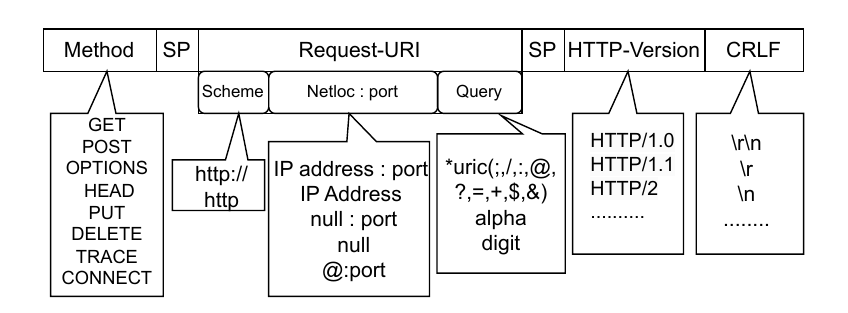}
   \caption{HTTP Request Packet Format with values.}
    \label{fig:httpget_param}
\end{figure}

The block diagram in Figure \ref{fig:httpget_param} shows the fields in the HTTP request packet shown in Figure \ref{fig:d3dHttpPkt}(a) that can be subject to mutation fuzzing. 
We have identified six broad fields to fuzz: \textit{method}, \textit{SP} (on left), \textit{Request-URI}, \textit{SP} (on right), \textit{HTTP-version} and \textit{CRLF}. \textit{Request-URI} is further divided into three subfields: \textit{scheme}, \textit{netloc:port} and \textit{query}. 
Thus, a total of eight fields are identified, whose values are fuzzed based on their seed values.  \textit{SP} is \textit{space} character.  The blocks below the respective fields show possible fuzzed values. A '*' indicates multiple values possible, \textit{alpha} and \textit{digit} indicate any alphabet and decimal digits, respectively.
For example, \textit{method} field can take any standard HTTP method, like \textit{GET}, \textit{POST}, \textit{OPTIONS}, \textit{HEAD}, \textit{TRACE}, \textit{CONNECT}, \textit{PUT}, or \textit{DELETE}, and \textit{scheme} in \textit{request-uri} can take \textit{http://} or \textit{http}. 


\begin{figure}
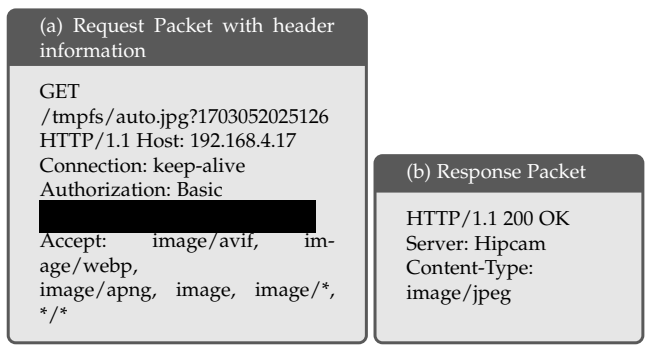

\centering
\begin{frame}

\resizebox{\linewidth}{!}{
\begin{tabular}{ccc}
\begin{tcolorbox}
[colback=gray!20!white,colframe=gray!70!black,width=6cm, title= (a) Request Packet with header information]
GET /tmpfs/auto.jpg?1703052025126  HTTP/1.1
Host:  192.168.4.17\\
Connection: keep-alive\\
Authorization: Basic\\\colorbox{black}{YWRtaW46YWRtaW4xMjM=}\\
Accept: image/avif, image/webp, \\ image/apng, image, image/*, */*
\end{tcolorbox}
\begin{tcolorbox}[colback=gray!20!white,colframe=gray!70!black, width=4.5cm,  title=(b) Response Packet]
HTTP/1.1 200 OK\\
Server:  Hipcam\\
Content-Type: image/jpeg\\
\end{tcolorbox}
\end{tabular}}
\end{frame}

\caption{HTTP Packets requesting image in D3D Camera.}
   \label{fig:d3dHttpPkt_imgext}
\end{figure}

Figure \ref{fig:d3dHttpPkt_imgext}(a) shows another HTTP packet that is carrying the GET request for live image from D3D IoT camera. This packet also contains \textit{authorization} tag, as in Figure \ref{fig:d3dHttpPkt}(a). Figure \ref{fig:d3dHttpPkt_imgext}(b) shows the corresponding response packet containing \textit{HTTP/1.1 200 OK}. 
Similar to that in Figure \ref{fig:d3dHttpPkt}(b), we have identified several fields, like \textit{Request-URI}, \textit{HTTP-version}, \textit{Host} and \textit{accept} tag ( reveals the image formats supported by the device) to apply mutation. 
Surprisingly, a large number of fuzzed values for  \textit{accept} tag can be generated, and each of these can trigger a valid response.  
Note that we do not fuzz any of the response packets because we aim not to inject any response packet into the network. A large volume of network traces of each IoT device is inspected to automatically extract the seed inputs, e.g., \textit{/web/cgi-bin/hi3510/ptzctrl.cgi?} is a seed in \textit{Request-URI} field, and then mutated inputs are generated for every such field. Thus, the field-specific fuzzed inputs lead to crafting structurally valid packets that can be injected into IUT.

\subsubsection{Fuzzing RTSP Packets}
Figure \ref{fig:d3dRTSPPkt}(a) and Figure \ref{fig:d3dRTSPPkt}(b) show a RTSP (Real Time Streaming Protocol \cite{RTSP_rfc2326}) request and corresponding response packet respectively, observed in D3D IoT camera.  
While HTTP packets often carry the control commands into the camera, RTSP packets deliver the actual live audio/video feed from the device to its controller. Importantly, these request packets do not contain any authentication data.

\begin{figure}[h]
\centering
\begin{frame}

\resizebox{\linewidth}{!}{
\begin{tabular}{ccc}
\begin{tcolorbox}
[colback=gray!20!white,colframe=gray!70!black,width=6cm, title=(a) Request Packet with RTSP method and URL.]
OPTIONS rtsp://192.168.4.6:554/  RTSP/1.0 \\
CSeq: 2\\
User-Agent: LibVLC/3.0.20 (LIVE555 Streaming Media v2016.11.28)\\
\end{tcolorbox}
\begin{tcolorbox}[colback=gray!20!white,colframe=gray!70!black, width=4.5cm,  title=(b) Response Packet with 200 OK.]
RTSP/1.0 200 OK\\
CSeq: 2\\
Public:  OPTIONS, DESCRIBE, PLAY, PAUSE, SETUP, TEARDOWN, SET\_PARAMETER, GET\_PARAMETER\\

\end{tcolorbox}
\end{tabular}}
\end{frame}

\caption{RTSP packets with URL in D3D Camera.}
   \label{fig:d3dRTSPPkt}
\end{figure}

\begin{figure}[h]
   \centering
   \includegraphics[width=\linewidth]{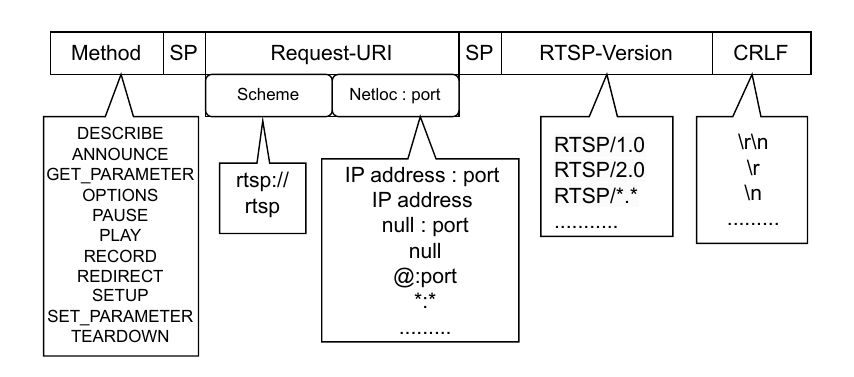}
   \caption{RTSP Packet Format with values.}
    \label{fig:rtsp_pktFormat}
\end{figure}

Similar to HTTP packets, we consider four distinguishable fields in a RTSP request packet: \textit{method}, \textit{request-URI}, \textit{RTSP-Version} and \textit{CRLF} (shown in Figure \ref{fig:rtsp_pktFormat}) in order to apply mutation.
For example, \textit{method} field can have values like \textit{DESCRIBE}, \textit{ANNOUNCE} and \textit{OPTIONS}, to name a few.
\textit{Request-URI} field is further divided into two subfields in this protocol based on differentiable strings: \textit{scheme} - indicates the name of the protocol, and \textit{netloc:port} - indicates the IP address of the device, and the transport port accepting RTSP packets. 
We extract a number of RTSP packets from the network traces of the IUT and extract a set of field-specific seed inputs.

\subsubsection{Fuzzing TP-Link SmartHome/JSON Packets}
\begin{figure}[h]
   \centering
   \includegraphics[width=0.8\linewidth]{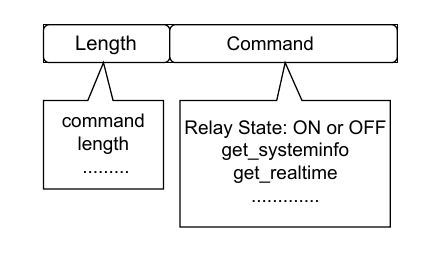}
   \caption{ TP-Link SmartHome/JSON Packets with values.}
    \label{fig:tplinksmarthome_pktsformat}
\end{figure}

Figure \ref{fig:tplink_cmd} (a) shows an example of a request packet in a proprietary protocol, called TP-Link SmartHome/JSON protocol \cite{TP-LInkSmartHomeProtocol} that is carrying a command to operate a TP-Link Kasa Smart Plug; the corresponding response packet is shown in Figure \ref{fig:tplink_cmd} (b). 
Similar to RTSP, this request packet does not contain any authentication data. 
The request packet has JSON data in its payload. We identify two fields (Figure \ref{fig:tplinksmarthome_pktsformat}) to apply mutation: \textit{length} - size in bytes of the command, and  \textit{command} -- exact JSON command. 

Table \ref{tab:cmd_smartplug} shows a set of encoded commands that may be present in the \textit{command} field (a complete list of such commands can be found in the protocol specification \cite{tplinkplugcmds}, \cite{tplinkplugkasacmds}). These commands and encoded values can be used to fuzz the smart plug. A deeper inspection reveals that the commands are specified in YAML. 

\begin{table*}[h]
\caption{Commands and encoded payloads associated with TP-Link Smart Plug used for Mutation Fuzzing.}
\label{tab:cmd_smartplug}
\resizebox{\linewidth}{!}
{
\begin{tabular}{|l|l|l|}
\hline
Command Type & Command with encoded payload   & Information    \\ \hline \hline
\multirow{4}{*}{\begin{tabular}[c]{@{}l@{}}System \\Commands\end{tabular}}    & \{"system":\{"get\_sysinfo":null\}\} \textcolor{blue}{\{AAAAI9Dw0qHYq9+61/XPtJS20bTAn+yV5o/hh+jK8J7rh+vLtpbr\} }                                     & \begin{tabular}[c]{@{}l@{}}Get System Information \\(e.g. Software \& Hardware \\Versions,MAC, deviceID, \\hwID etc.)\end{tabular} \\ \cline{2-3} 
                                     & \{"system":\{"set\_relay\_state":\{"state":1\}\}\}   \textcolor{blue}{\{AAAAKtDygfiL/5r31e+UtsWg1Iv5nPCR6LfEsNGlwOLYo4HyhueT9tTu36Lfog==\}}                     & Turn On    \\ \cline{2-3} 
                                     & \{"system":\{"set\_relay\_state":\{"state":0\}\}\}   \textcolor{blue}{\{AAAAKtDygfiL/5r31e+UtsWg1Iv5nPCR6LfEsNGlwOLYo4HyhueT9tTu3qPeow==\}}                     & Turn Off   \\ \cline{2-3} 
                                     & \{"emeter":\{ "get\_realtime":null \}\}  \textcolor{blue}{\{AAAAJNDw0rfav8uu3P7Ev5+92r/LlOaD4o76k/6buYPtmPSYuMXlmA==  \}}                  &  \begin{tabular}[c]{@{}l@{}}Get Realtime Current\\ and Voltage Reading \end{tabular}    \\ \cline{2-3} 
                                     
                                    \hline                
\end{tabular}}
\end{table*}

\subsection{Vulnerability Classification}
\label{sec:vul_iot}
Based on the semantics of the crafted packets with mutated inputs and the applicability on the IoT devices, we categorize the vulnerabilities reported in this paper as follows.

\begin{itemize}
    \item \textbf{IoT Access Credential Leakage (D3D\_000):} 
    To check this vulnerability, the tool takes a sample of network traffic and looks for those HTTP packets that carry \textit{authorization} tag. 
    A successful discovery of this vulnerability is achieved only when the authentication scheme is \textit{Basic} as it indicates an encoded, e.g., Base-64, and not an encrypted credential of the IoT device (fitting in D3D IoT Camera). 
    
    This vulnerability can be exploited by crafting and then sending a HTTP packet mutated \textit{GET} or \textit{PUT} request having a command to the IUT. If the device generates a valid HTTP response, like status code as \textit{200 OK} and content as \textit{set ok.} (as shown in Figure \ref{fig:d3dHttpPkt}(b)), then we consider that the vulnerability is exploited successfully. 

    \item \textbf{Creep IoT Live Image (D3D\_002):} 
    To test this vulnerability, our tool at first crafts a set of specific HTTP packets having \textit{method} as \textit{GET} and \textit{Request-URI} as mutated values indicating paths to certain image files within the directory structure of the IUT. If any of such packets trigger a valid response such as \enquote{HTTP/1.1 200 OK} having \textit{content-type} as \enquote{image/jpeg}, then the device has this vulnerability (fitting in D3D IoT Camera). 
    
    This vulnerability can be exploited successfully if the attacker is able to extract the actual image from the subsequent network packets.

    \item \textbf{IoT Command Injection (D3D\_003)}
    To test this vulnerability, the tool crafts a set of HTTP packets with \textit{method} as \textit{GET} and \textit{query} in \textit{Request-URI} field as mutated values, e.g., \textit{-a=ledt}, and then sends it to the IUT. Note that such an HTTP packet requires \textit{authorization} tag with an appropriate value.  
    If any of such packets trigger a valid HTTP response, like status code as \textit{200 OK} having content as \textit{set ok}, then the device has this vulnerability. In this case, the device actually executes the command and hence leads to a successful exploit. 
    
    \item \textbf{Creep Live IoT Image (Ezviz\_001):} 
    This vulnerability is similar to D3D\_002. We consider it as a separate one as the directory structure in Ezviz IoT camera is not same as D3D, leading to different seed and mutated inputs. 

    \item \textbf{Sneak IoT Live Video Stream (D3D\_001):} 
    To test this vulnerability, the tools at first craft a set of RTSP request packets having fuzzed values in its \textit{method}, \textit{netloc:port} in \textit{Request-URI} and \textit{CSeq} fields, and then sends it to the IUT. 
    If any of such packets trigger a valid response, e.g., having status code \enquote{200 OK} and the same \enquote{CSeq} value, then the device has this vulnerability. 
    
    If the attacker can view the camera feed, e.g., using VLC media player \cite{VLC} or iSpy \cite{iSpy} or Cameradar \cite{Cameradar}, the vulnerability stands exploited successfully.

    \item \textbf{Sneak IP camera Live Video Stream (Ezviz\_000):} 
    This vulnerability is similar to D3D\_001. 
    Again, it is a separate payload in the Ezviz IoT camera, and it is not the same as D3D, leading to different seed and mutated inputs. 
    
    \item \textbf{IoT Command Injection (TP-Link\_Kasa\_000):}
    To test this vulnerability, the tool crafts a set of \textit{TP-Link SmartHome/JSON} packets having any one of the fuzzed values (e.g., ON or OFF, reboot, delay, get\_systeminfo, etc.) in \textit{Cmd} tag along with its length and then sends it to the IUT. 
    If any of such packets trigger a valid response, e.g., having \textit{error\_code} as 0, then we consider it as a successful discovery of this vulnerability. Note here that these protocol packets do not contain any authentication mechanism. Importantly, such a device can accept any packet, irrespective of its current status, like On or Off. 
    Successful exploitation is automatically achieved when a valid response is received. 
\end{itemize}

\begin{table*}[]
    \centering
\caption{Summary of vulnerabilities in IoT devices (IP cameras: D3D, Ezviz  and Smart Plug: TP-Link Kasa)}
\label{tab:explo_vulnIoT}
\begin{tabular}{|l|c|c|c|c|c|c|c|}\hline
\diagbox[width=15em]{IoT\\Devices}{Vulnerability(Device\_\#)}&
  D3D\_000 & D3D\_001 & D3D\_002 & D3D\_003 & Ezviz\_000 & Ezviz\_001 & TP-Link\_Kasa\_000\\ \hline \hline
D3D Camera& \cmark & \cmark & \cmark  & \cmark & \xmark  & \xmark & \xmark \\ \hline
Ezviz Camera& \xmark &  \xmark  &  \xmark  &  \xmark & \cmark &\cmark &\xmark \\ \hline
TP-Link Kasa Smart Plug& \xmark &  \xmark  &  \xmark  &  \xmark & \xmark & \xmark & \cmark\\ \hline

\end{tabular}
\end{table*}

\section{IoTFuzzSentry on Commercial IoT Devices}
\label{sec:integratoin-into-cotopaxi}

\subsection{Implementation of IoTFuzzSentry}

\begin{figure}[h]
\includegraphics[width=\linewidth]{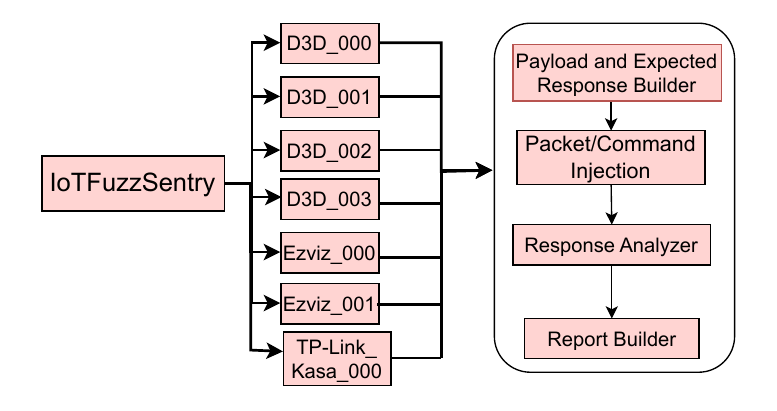}
  \caption{IoTFuzzSentry Structure.}
  \label{fig:iotfuzzsentry_structure}
\end{figure}

Figure \ref{fig:iotfuzzsentry_structure} shows the moduler code structure of IoTFuzzSentry, the entire code base is in a Python file named as \textit{IoTFuzzSentry.py}. 
The Python file contains a total of seven Python3 classes, one for each of seven newly classified vulnerabilities, namely \textit{D3D\_000}, \textit{D3D\_001}, \textit{D3D\_002}, \textit{D3D\_003}, \textit{Ezviz\_000}, \textit{Ezviz\_001} and \textit{TPLink\_Kasa\_000}. The summary of vulnerabilities is shown in Table \ref{tab:explo_vulnIoT} 
Each of these classes has a common functional structure and control flow. The first function builds the payloads and the corresponding expected responses, the second function crafts protocol-specific packets, the third function analyses the responses, and the third function produces a comprehensive report.

\subsection{Discovering Vulnerabilities using IoTFuzzSentry}
\subsubsection{D3D\_000}
To test this vulnerability, we have considered as input a sample network traffic of a D3D IoT camera containing 20543 packets. 
This vulnerability is discovered passively by searching for HTTP request packets that have \textit{authorization} tag with the scheme as \textit{BASIC}. 
The tool reports 394 such packets in the sample. Note that the credential revealed in this tag is not necessary to be valid; to check for validity, the credential needs to be encapsulated in a request packet and seen if a valid response is received.

\subsubsection{D3D\_001 and Ezviz\_000}
In the D3D camera, we have found \textit{six} seed URIs from the RTSP packets in the legitimate traffic. 
Our mutation fuzzer then generates 200 mutated URIs, using a similar technique as in \cite{fuzzingbook2024}. 
Among these mutated URIs, the camera generates valid responses to 30 URIs, and hence, we discover the D3D IoT camera suffers D3D\_001. 
Some examples of valid mutated RTSP URIs in this camera are \textit{DESCRIBE rtsp://192.168.4.17:554/}, \textit{PLAY rtsp://192.168.4.17/}, and \textit{rtsp://\{\}:\{\}/ }.

Similarly, in EZVIZ\_000, we have got 28 out of 200 valid mutated RTSP URIs. Some examples of seed inputs are \textit{rtsp://192.168.4.7/0} and \textit{rtsp://192.168.4.7/1/stream1}. Some example of valid mutated URIs are \textit{rtsp://192.168.4.7/1} corresponding to first seed and \textit{\url{rtsp://192.168.4.7/stream2}}
corresponding to the second seed.

\subsubsection{D3D\_002}
In D3D\_002, we have extracted seed URIs like \textit{http://192.168.4.18/
tmpfs/auto.jpg?1703052025126} from HTTP request packets, which shows the temporary folder name \textit{tmpfs} and the image file as \textit{auto.jpg}. 
Out of 200 mutated URIs, 28 are found valid in this case. 
Some examples of valid mutated URIs are \textit{http://192.168.4.18/tmpfs/auto.jpg?17030h5202
og5126}, \textit{\url{http://192.168.4.18/tVmp/auto.jpg?13 30202514}}, and \textit{(\url{http://192.168.4.18/mtmPfs/auto.jpg?703024126})}

\subsubsection{D3D\_003}
\begin{figure}
\centering
\subfloat[Movement at 10:23:09 AM]{\label{4figs-a} \includegraphics[width=0.48\linewidth]{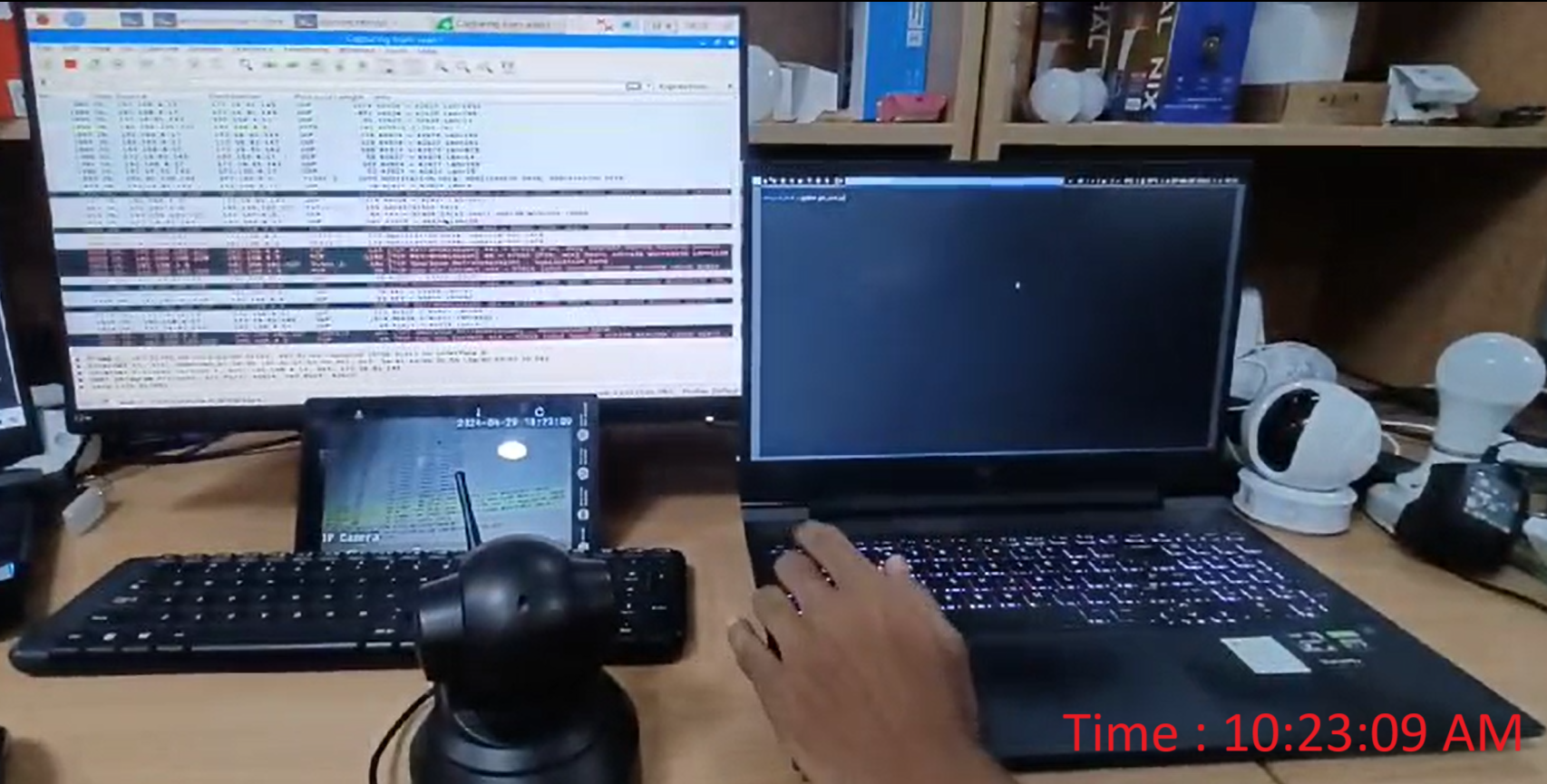}}
\hfill
\subfloat[Movement at 10:23:10 AM]{\label{4figs-b} \includegraphics[width=0.48\linewidth]{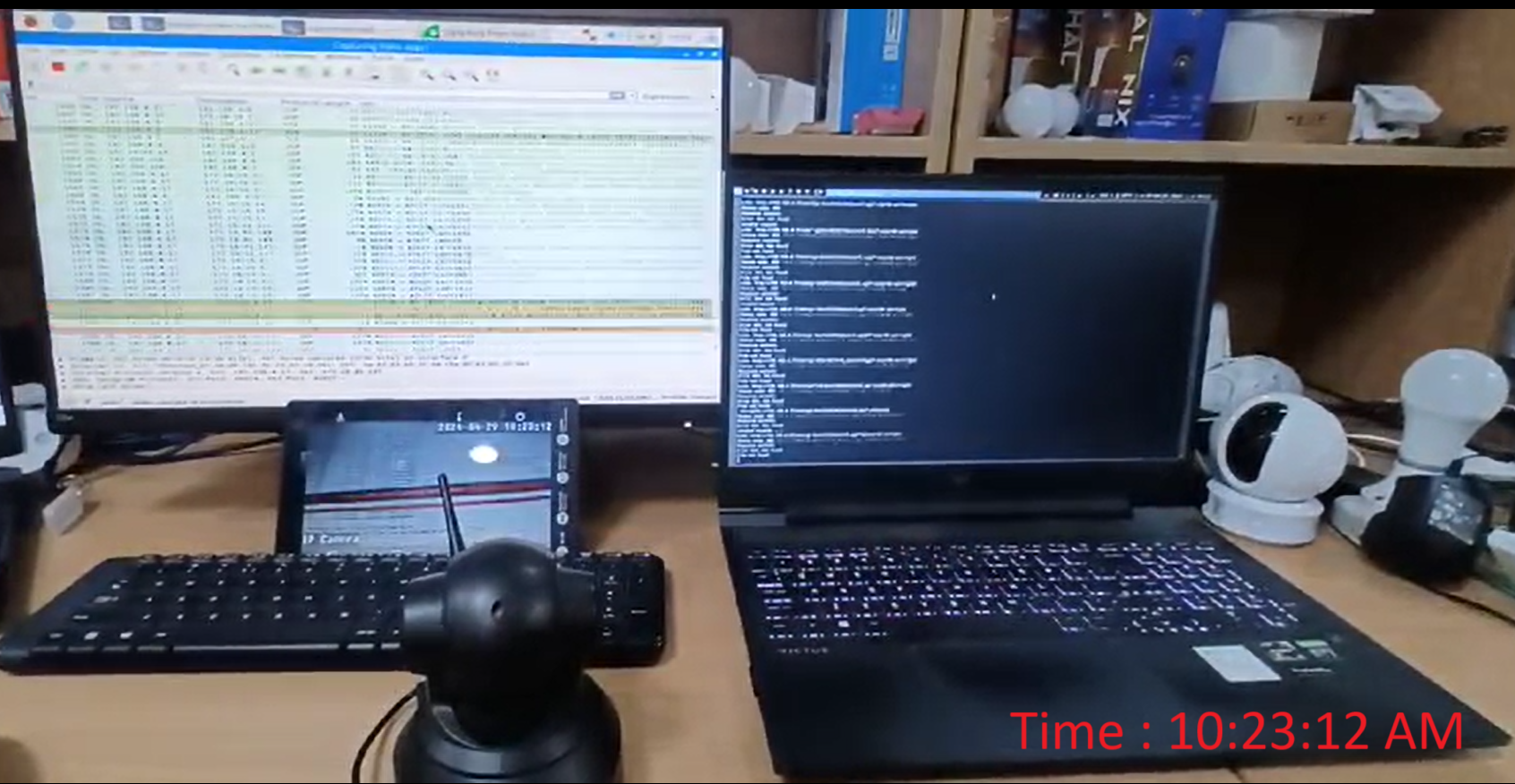}}%
\hfill
\subfloat[Movement at 10:23:16 AM]{\label{4figs-c} \includegraphics[width=0.48\linewidth]{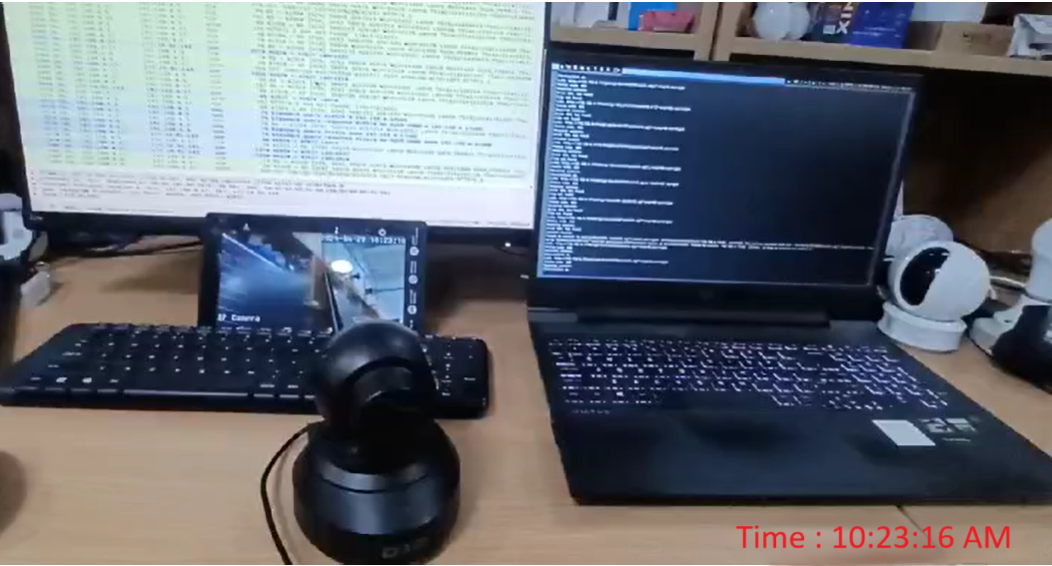}}%
\hfill
\subfloat[Movement at 10:23:27 AM]{\label{4figs-d} \includegraphics[width=0.48\linewidth]{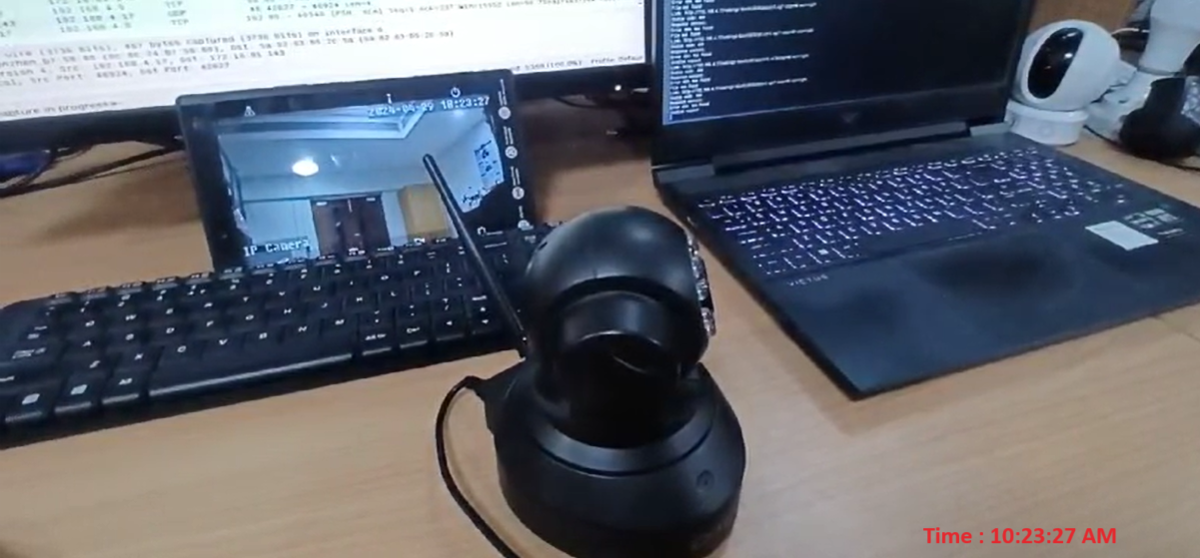}}%
\caption{Movement of D3D camera by exploiting vulnerabilities D3D\_000, D3D\_001, D3D\_002.}
   \label{fig:explo_d3d}
\end{figure}

Using a valid tag-value pair of \textit{Authorization: 
Basic 
\colorbox{black}{YWRtaW46Yml0c0AxMjM=}}, we have crafted HTTP request packets having mutated values in \textit{URI} tag. Out of 200, 39 mutated URIs turn out to be valid. An example of seed URI is \textit{http://192.168.4.18/web/cgi-bin/hi3510/ptzctrl.cgi?-step=0\&-act=right}, and a corresponding valid mutated URI is \textit{http://192.168.4.18/webA/cgi-(bin/hi3510/ptzctrl. cgi?-step=0\&-act=right}. Interestingly, each of the valid URIs has caused the camera to perform the corresponding action, like rotating the focus area to left or right with a status code as 200 (OK) in response.
Figure \ref{fig:explo_d3d} shows the camera positions from when such a packet is injected to when the camera settles down its focus. Figure \ref{fig:explo_d3d}(a) shows the camera focus at 10:23:09 AM, i.e., just before it starts rotation, Figure \ref{fig:explo_d3d}(b) at 10:23:10 AM, i.e., it started rotating to its right, Figure \ref{fig:explo_d3d}(c) at 10:23:16 AM, i.e, it started capturing an image after a little more rotation. The controller app is displaying the table image, and lastly \ref{fig:explo_d3d}(d) at 10:23:27 AM, i.e., it completes the rotation by focusing to its right direction; the images are clicked roughly 7-8 Seconds apart. 

\subsubsection{TP-Link\_Kasa\_000}
\begin{figure}[h]
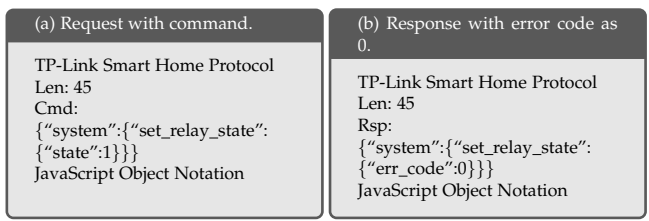

\centering
\begin{frame}

\resizebox{\linewidth}{!}{
\begin{tabular}{ccc}
\begin{tcolorbox}
[colback=gray!20!white,colframe=gray!70!black,width=6cm, title=(a) Request with command.]
TP-Link Smart Home Protocol\\
     Len: 45\\
     Cmd: \{“system”:\{“set\_relay\_state”:\\\{“state”:1\}\}\}\\
   JavaScript Object Notation\\
\end{tcolorbox}
\begin{tcolorbox}[colback=gray!20!white,colframe=gray!70!black, width=6cm,  title=(b) Response with error code as 0.]
TP-Link Smart Home Protocol\\
     Len: 45 \\
     Rsp: \{“system”:\{“set\_relay\_state”:\\\{“err\_code”:0\}\}\}\\
  JavaScript Object Notation

\end{tcolorbox}
\end{tabular}}
\end{frame}

\caption{TP-Link SmartHome/JSON packets with Command in TP-Link Smart Plug.}
   \label{fig:tplink_cmd}
\end{figure}

We have extracted seed payloads in TP-Link SmartHome/JSON protocol, e.g., \textit{AAAAKtDygfiL/5r31e+UtsWg1Iv5nPCR6L
fEsNGlwOLYo4Hyhue
T9tTu36Lfog==} that corresponds to Switch On and \textit{AAAAKtDygfiL/5r31e+UtsWg1Iv5nPCR6LfEsNGlw
OLYo4HyhueT9tTu3qPeow==} to Switch Off the plug.
Our mutation fuzzer generates mutated payloads, like \textit{\seqsplit{AAAAKtDygfiL/5r31e+UtsWg1Iv5nPC} \seqsplit{R6LfEsNGlwOLYo4HyhueT9tTu36Lfog==}}, which is then placed in a TCP packet with destination port as 9999 to send it to the plug. 
Figure \ref{fig:tplink_cmd} (a) shows a content, i.e., relay\_state = 1 (ON), in the request packet sent to the plug, and Figure \ref{fig:tplink_cmd} (b) shows the content, i.e., err\_code = 0, in the corresponding response packet in TP-Link SmartHome /JSON protocol. The response packet indicates that the mutated command has been executed without any error in the IoT plug. 

\subsection{Integration of IoTFuzzSentry into Cotopaxi}
In order to enhance its applicability and reachability, we have considered Cotopaxi as one of the possible Python-based open-sourced IoT fuzz testing tools for integrating IoTFuzzSentry.
 
\subsubsection{Existing Capabilities in Cotopaxi} 
Released as open-sourced in 2019, Cotopaxi has become one of the popular IoT security testing toolkits capable of fuzzing 14 IoT protocols, namely AMQP, CoAP, DTLS, HTCPCP, HTTP, HTTP/2, gRPC, KNX, mDNS, MQTT, MQTT-SN, QUIC, RTSP, and SSDP, including DTLS that works on top of UDP. 
This tool has been built to craft protocol-specific packets and analyze the response of those packets from the targeted IoT devices. 
Typically, Cotopaxi requires as input a range of port numbers on which security testing needs to be performed. 
For example, given a range of port numbers, \textit{service ping} can test any 14 application layer protocols in an IUT. 

At its core, Cotopaxi has ten high-level packages that offer various services, like \textit{service ping}, \textit{server fingerprinting}, \textit{device identification}, and \textit{vulnerability testing}. 
For example, it can actively reconnaissance an IoT device, called \textit{resource\_listing}, where it can discover the directory structure in the device. 
This command takes a specific resource, e.g., \enquote{cotopaxi/lists/urls/coap urls.txt}, as input and checks if such a file exists in the directory structure of IUT; typically, such a file name is placed in an HTTP GET request using \textit{URI} tag.

\begin{figure}[h]
\includegraphics[width=\linewidth]{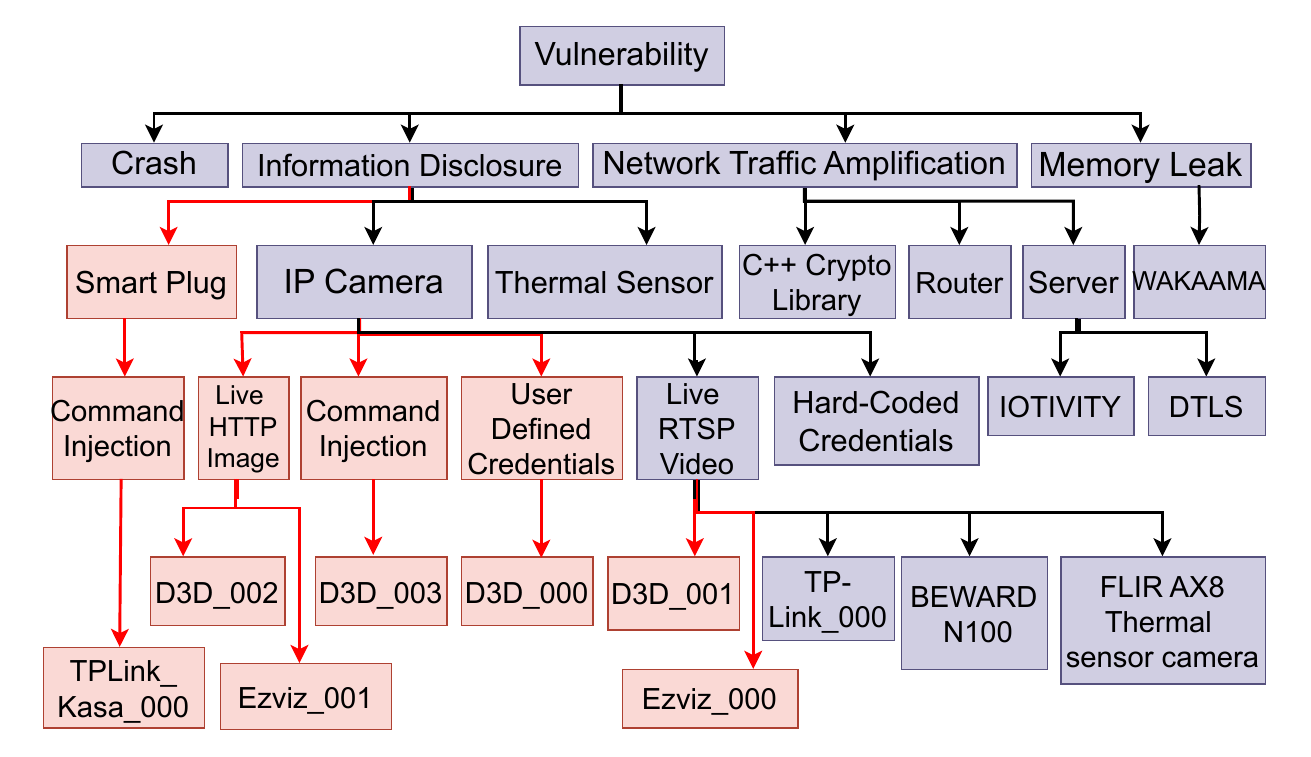}
  \caption{Our classification of vulnerability packages in Cotopaxi. Packages indicated by grey and pink boxes denote the currently available and our new contributions respectively.} 
  \label{fig:coto_vul_category}
\end{figure}

Among other services, vulnerability testing can check if certain known vulnerabilities, i.e., those registered in the CVE database, are present in an IoT device. Currently, Cotopaxi supports 14 such vulnerabilities, that we broadly categorize into four types based on the impact of the test being conducted (Figure \ref{fig:coto_vul_category}): \textit{crash}, \textit{information disclosure}, \textit{network traffic amplification} and \textit{memory leak}. 
Because our work in this paper focuses on privacy leakage, we focus primarily on the \textit{information disclosure} category.

The information disclosure category is further categorized based on device types into IoT Cameras and IoT Thermal Sensors. 
Further, based on application protocols and the type of data being leaked, the vulnerabilities in IoT cameras are further classified into RTSP video and hard-coded access credentials. Finally, based on the exact models, three vulnerabilities in the category of live RTSP video can be tested: TP-Link\_000, BEWARD N100, and FLIR AX8 Thermal Sensor Camera. 
A specific Cotopaxi command takes as input an argument, like \textit{RTSP --vuln TP-Link\_000}, to test a particular known vulnerability. 
Note that an existing payload, e.g., in RTSP, that can test vulnerability in the BEWARD N100 camera may not be applicable to other cameras due to differences in field-specific values. 

\subsubsection{Results of Testing Existing Vulnerabilities}
We report the results of testing already supported vulnerabilities in Cotopaxi on the three IoT devices used in this paper. 
Testing the vulnerability of HTTP-based hard-coded credentials leakage, i.e., \textit{--vuln FOSCAM\_001 --verbose}, resulted in \textit{Bad Request} or \textit{probably not vulnerable} when applied on D3D and Ezviz IoT cameras. 
This indicates that the existing fuzzed payloads are suitable only in Foscam IoT cameras.

Similarly, testing of RTSP-based denial-of-service vulnerability, i.e., \textit{FOSCAM\_000}, on D3D and Ezviz resulted in \textit{Bad Request} or \textit{probably not vulnerable}. This indicates that the existing fuzzed RTSP payloads are suitable only in Foscam,  and possibly on BEWARD N100, TP-Link TL-SC3130, and FLIR AX8 thermal sensor IoT cameras. 

Finally, the testing of existing HTTP-based live video stream access vulnerability, i.e., \textit{BEWARD\_000}, on either camera showed that the object, i.e., the specified image or audio or video files, \textit{Not Found} or the device is \textit{probably not vulnerable}. Our analysis shows that the fuzzed HTTP GET request sent by Cotopaxi is accepted by the device (indicated by \textit{HTTP GET: SUCCESS} in its report), yet the reports indicate that these devices do not suffer this vulnerability.

\subsubsection{Merging IoTFuzzSentry into Master Code Base in Github}

\begin{figure}[h]
\centering
\includegraphics[width=\linewidth]{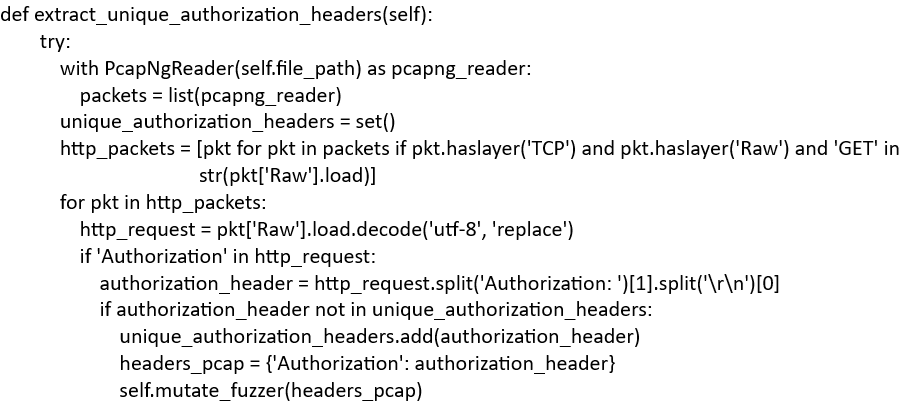}
  \caption{Processing Code for credentials leakage.}
  \label{fig:coto_integ_code1}
\end{figure}

\begin{figure}[h]
\centering
\includegraphics[width=\linewidth]{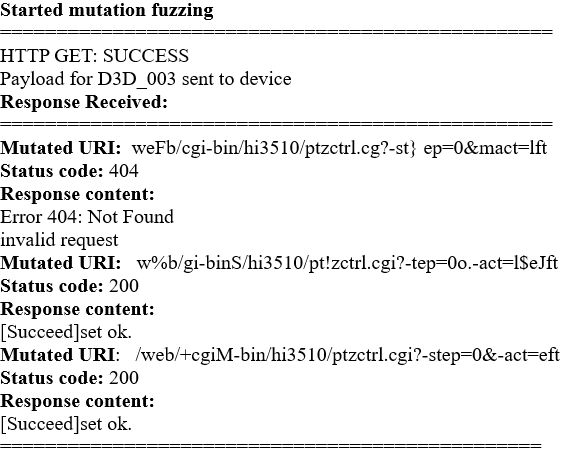}
 \caption{Mutated URIs for D3D\_003 vulnerability.}
\label{fig:resultURI_d3d003}
\end{figure}

\begin{figure}[h]
\centering
\includegraphics[width=\linewidth]{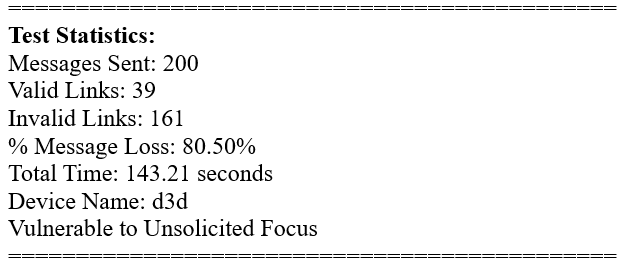}
 \caption{Report for D3D\_003 vulnerability.}
\label{fig:report_d3d003}
\end{figure}

After completing a successful test in our local copy, we have uploaded the complete Python module of IoTFuzzSentry into the master repository of Cotopaxi in Github \cite{our-contrib-cotopaxi-github}.
The new \textit{pull request} in Github shows \textit{verified}, which indicates that the new code has passed the necessary quality checks.
Figure \ref{fig:coto_integ_code1} shows a code sample used to extract \textit{authorization} tag and its value to test D3D\_000. 
We ensure that our new modules are executed using the same command structure as in existing Cotopaxi. For example, to test D3D\_000 and D3D\_003, the commands are \textit{python3 -m cotopaxi.iotfuzzsentry 192.168.4.17 --protocol HTTP --vuln D3D\_000 --verbose} and \textit{python3 -m cotopaxi.iotfuzzsentry 192.168.4.17 554 --vuln D3D\_003 --verbose} respectively. 
In the \textit{verbose} mode, the tool generates the details of mutated inputs (URIs in this case) along with the status code in its response (an example is shown in Figure \ref{fig:resultURI_d3d003}). Irrespective of \textit{verbose} mode present or not, the tool generates a detailed report as shown in Figure \ref{fig:report_d3d003} showing the number of mutated packets send, valid and invalid responses received, message loss (\%) and the time required to complete the test.

In the core, Cotopaxi contains \textit{vulnerability\_tester} module that interacts with a database, called \textit{Vulnerability}, that contains raw payloads (one in each .raw files) and a configuration file (called vulnerabilities.yaml). 
An entry in the .yaml file corresponds to one vulnerability, and it is separated from another using the '-' symbol. 
Hence, we have added seven entries in this file to accommodate seven new vulnerabilities. An example of such an entry in a .yaml file is shown below, corresponding to D3D\_000 shown in Figure \ref{fig:vulD3D000_yaml}.

\begin{figure}[h]
\centering
\includegraphics[width=\linewidth]{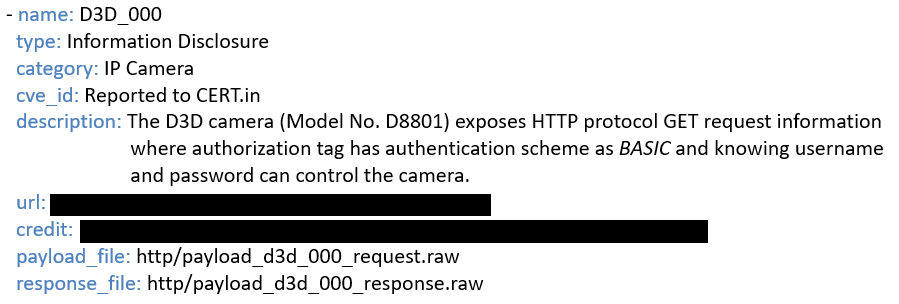}
  \caption{Vulnerability D3D\_000 in .yaml file}
  \label{fig:vulD3D000_yaml}
\end{figure}

\section{Effectiveness of IoTFuzzSentry}
\label{sec:effect_fuzzer}
Typically, the effectiveness of a fuzzing tool, like libFuzzer \cite{Libfuzzer} and AFL++ \cite{fioraldi2020afl++}, can be assessed by using \textit{coverage}, e.g., \textit{code coverage} or \textit{path coverage}. The code under test, in this case, is made available to the fuzz tester, where the locations of code, like statement number or function name, are identified for each of the test cases executed. The more unique statements, the higher the coverage. 
In our case, the code executing in an IoT device is not available, and hence, we perform \textit{black-box} fuzzing. 
However, it may respond with a packet to an injected packet. For example, if an HTTP packet is sent to an IoT device having TCP port 80 activated, then it can produce either a valid or an invalid HTTP response. 
Therefore, we extend the concept of \textit{coverage} to what we call as \textit{response coverage} (similar to AFLNet \cite{AFLNET2020IEEE} and Snipuzz \cite{feng2021snipuzzACM}) to assess the efficacy of IoTFuzzSentry.

\subsection{Response Coverage}
Table \ref{tab:resp_coverage} shows the response coverage of IoTFuzzSentry in both D3D and Ezviz IoT cameras using both HTTP and RTSP fuzzed packets. 
Testing D3D\_001, 48\%, and 52\% of packets with mutated payloads produce valid (i.e., 200 OK) and invalid (i.e., 404 Not Found) responses, respectively. 
The packets causing other responses, like 400 (Bad Request), are not considered because the payloads resulted in malformed HTTP packets (20\%). 
Similarly, about 32\% of HTTP packets have resulted in valid responses from the Ezviz camera while testing Ezviz\_001. 

\begin{table}[h]
\caption{Response Coverage using RTSP responses for live video feed and HTTP responses for live images.}
\label{tab:resp_coverage}
\resizebox{\linewidth}{!}{
\begin{tabular}{|l|cc|cc|} \hline
\diagbox[width=15em]{IPCam}{Response Coverage} & \multicolumn{2}{c|}{\begin{tabular}[c]{@{}c@{}}RTSP URL Responses \\ for Live Video Streaming\end{tabular}} & \multicolumn{2}{c|}{\begin{tabular}[c]{@{}c@{}}HTTP URL Responses \\ for Live Image Extraction\end{tabular}} \\ \cline{2-5}                                                      & \multicolumn{1}{c|}{Valid}                        & \multicolumn{1}{c|}{Invalid}                       & \multicolumn{1}{c|}{Valid}                        & \multicolumn{1}{c|}{Invalid}                        \\ \hline \hline
D3D Camera  & \multicolumn{1}{c|}{48\%}                         & 52\%                                               & \multicolumn{1}{c|}{44\%}                         & 55\%                                                \\ \hline
Ezviz Camera  & \multicolumn{1}{c|}{35\%}                         & 65\%                                               & \multicolumn{1}{c|}{32\%}                         & 68\%                                                \\ \hline
\end{tabular}}
\end{table}

\subsection{Responsible Disclosure of Vulnerabilities}
We have reported via email all the discovered vulnerabilities to the respective vendors and CERT \cite{certin} in India. 
So far, we have received two CVE numbers CVE-2024-41623 \cite{D3D_CVE-2024-41623} and CVE-2024-42531 \cite{EzCVE-2024-42531} 
The vendor of TP-Link Kasa Plug has requested some additional time so that they can patch it before we publish the CVE. 

\subsection{Discussion on Reasons of the Vulnerabilities}
\subsubsection{Lack of integrity check}
Our analysis of the vulnerabilities' causes reveals that these devices lack simple integrity checks in the application protocols. This is evident from the fact that each device accepts mutated payloads in the protocol packets. 

\subsubsection{Lack of mutual authentication}
In D3D and Ezviz, the cameras accept unauthenticated requests from any host and allow access to its live images or videos. 
This indicates that the devices lacks basic authentication check in such requests packets. Absence of encryption in the application protocols in IoT devices may not be a new revelation \cite{cynthia2019Springer, Nebnione2020MDPIFI, abbasi2022IEEEAccess} and it is not an exception in case of these two devices.

\subsubsection{Unsolicited Leakage of Server Communication}
Finally, our analysis shows that certain mutated payloads force the IoT device to reveal protocol-specific content while communicating with its cloud server, which can help social engineering about the device. For example, a HTML page with embedded JavaScript can reveal the functions related to language selection, AJAX requests (\textit{CreateRequest(), HttpRequest(), SendCGICMD()}) and \textit{on\_load()}) function. These functions actually calculates the current date and time which then send to server using CGI commands. 

\begin{table*}[h]
\caption{Comparison of IoTFuzzSentry with state-of-the-art for executing payloads of respective tools in our tool.}
\label{tab:comp_stateofarttools}
\resizebox{\linewidth}{!}{
\begin{tabular}{|c|c|c|c|cc|l|l|l|}
\hline
\multirow{2}{*}{Fuzzers} & \multirow{2}{*}{\begin{tabular}[c]{@{}c@{}}Types of \\ Fuzzer\end{tabular}} & \multirow{2}{*}{\begin{tabular}[c]{@{}c@{}}Target\\ System/Firmware\end{tabular}}                                                  & \multirow{2}{*}{\begin{tabular}[c]{@{}c@{}}Seed\\ Input\end{tabular}}                                                  & \multicolumn{2}{c|}{Payloads}        & \multicolumn{1}{c|}{\multirow{2}{*}{\begin{tabular}[c]{@{}c@{}}Execution Summary on our \\ IoT Devices\end{tabular}}}    & \multicolumn{1}{c|}{\multirow{2}{*}{\begin{tabular}[c]{@{}c@{}}Vulnerability\\ Types\end{tabular}}}                                & \multirow{2}{*}{\begin{tabular}[c]{@{}c@{}}Feedback\\ System\end{tabular}}       \\ \cline{5-6}
                         &  &  && \multicolumn{1}{c|}{HTTP}     & RTSP & \multicolumn{1}{c|}{}   & \multicolumn{1}{c|}{}     &      \\ \hline \hline
Fuzzoswki  \cite{Fuzzowski}              & Blackbox                                                                    & \begin{tabular}[c]{@{}c@{}}Network Protocol \\ (LPD, IPP, \\ BACKNet, \\ Mod-Bus, TFTP, \\ HTTP)\end{tabular}                      & \begin{tabular}[c]{@{}c@{}}Packet\\ Payload\end{tabular}                                                               & \multicolumn{1}{c|}{1 (POST)} & NA   & \begin{tabular}[c]{@{}l@{}}Response : "Received" \\ or no information\end{tabular}  & 1. Unknown crash   & \begin{tabular}[c]{@{}l@{}}No Feedback\\ System\end{tabular}                     \\ \hline
AFLNet  \cite{AFLNET2020IEEE}                  & Graybox                                                                     & \begin{tabular}[c]{@{}c@{}}Media streaming server \\ (e.g. LIVE555)\\ RTSP, FTP, DTLS12, \\ DICOM, SMTP, \\ DAAP-HTTP\end{tabular} & \begin{tabular}[c]{@{}c@{}}Actual message \\ sequence of\\ protocol packets\end{tabular}                               & \multicolumn{1}{c|}{4 (POST)} & 29   & \begin{tabular}[c]{@{}l@{}}1. Response :401  \\ Unauthorized\\ 2.  Remote connection\\  opened, sent RTSP \\ payloads repeatedly\\ to clients and received \\ response as closes streams\\ 3. Response as Program\\  Aborted with location\end{tabular} & \begin{tabular}[c]{@{}l@{}}1. Memory Leak\\ 2. Stack Buffer Overflow\\ 3. Segmentation Violation\\ 4. Unknown Crashes\end{tabular} & \begin{tabular}[c]{@{}l@{}}1. Coverage\\ 2. Response \\ from SUT\end{tabular}    \\ \hline
Snipuzz  \cite{feng2021snipuzzACM} & Blackbox & IoT Firmware  & \begin{tabular}[c]{@{}c@{}}API program as\\ Message payloads\end{tabular}                                              & \multicolumn{1}{c|}{-}        & -    & --                                      & \begin{tabular}[c]{@{}l@{}}1. Null Pointer Exceptions\\ 2. Denial of Service\\ 3. Unknown Crashes\end{tabular}                     & \begin{tabular}[c]{@{}l@{}}1. Coverage\\ 2. Response \\ from server\end{tabular} \\ \hline
Doona  \cite{Doona}                  & Blackbox                                                                    & \begin{tabular}[c]{@{}c@{}}Network protocol\\ (HTTP, RTSP, \\ TFTP, \\ PROXY, POP,)\end{tabular}                                   & \begin{tabular}[c]{@{}c@{}}Test cases need to\\ be predefined as \\ per device and \\ protocol under test\end{tabular} & \multicolumn{1}{c|}{39}       & 65   & \begin{tabular}[c]{@{}l@{}}1. Response : Max \\ Requests (3) completed,\\  index: 3\\ 2. Response: 400 Bad \\ Request for RTSP payload\end{tabular}                                                                                                     & \begin{tabular}[c]{@{}l@{}}1. Buffer Overflow\\ 2. Format Strings Bugs\end{tabular}                                                & \begin{tabular}[c]{@{}l@{}}No Feedback \\ System\end{tabular}                    \\ \hline
\end{tabular}}
\end{table*}

\subsection{Impact of Disclosed Vulnerabilities}
To assess the impact of the discovered vulnerabilities, we have investigated the usage statistics, e.g., the number of downloads, in the Google Play Store of the companion mobile applications, i.e., the controller, and the user ratings of each of the three IoT devices used in our experiments. It reveals that the controller of D3D camera (i.e., iMegaCam or D3DCam), Ezviz camera (i.e., Ezviz App) and TP-Link Kasa smart plug (i.e, Kasa App) have been downloaded more than 10,000,  10 Million and 5 Million times respectively. 
There are about 1200, 224k and 120k users feedbacks in D3D camera, Ezviz Camera and TP-Link SmartPlug respectively, and the feedbacks are mainly related to the usage convenience and the device functionalities; the average ratings have been 3.7, 3.7, 4.6 and 4.7 stars respectively on a scale of 1 (low) to 5 (high) as on June, 2024. 
Based on the last user feedback, we argue in favor that our discovery of vulnerabilities can have a direct impact on the user community and cyber security enthusiasts.

To verify further the current state of deployments, we have searched for these devices in a well-known IoT search engine, \textit{Shodan} \cite{shodan}. It turns out that all three devices, i.e., D3D, Ezviz, and TP-Link SmartPlug, have an online presence across the globe. For example, currently, more than 500 D3D cameras are active online, and the search results show that all these cameras have HTTP daemon activated on several ports other than the standard TCP port 80. 
Thus, we believe that IoTFuzzSentry can be applied to any of these online devices with minor or no modification to its current code base, which we consider as part of our future work.

\subsection{Generalizability of IoTFuzzSentry Across IoT Devices}
\label{sec:iotfuzzsentry-6iotdevices}

\begin{table*}[h]
\centering
\caption{Summary of vulnerabilities in other IoT devices}
\label{tab:gen_6iotdevices}
\resizebox{\linewidth}{!}{\begin{tabular}{lllllll}
\hline
\multicolumn{1}{|c|}{\textbf{\begin{tabular}[c]{@{}c@{}}IoT\\ Device\end{tabular}}} & \multicolumn{1}{c|}{\textbf{\begin{tabular}[c]{@{}c@{}}Observed \\ Ports\end{tabular}}} & \multicolumn{1}{c|}{\textbf{\begin{tabular}[c]{@{}c@{}}Authentication \\ Scheme\end{tabular}}} & \multicolumn{1}{c|}{\textbf{\begin{tabular}[c]{@{}c@{}}Method\\ Used\end{tabular}}} & \multicolumn{1}{c|}{\textbf{\begin{tabular}[c]{@{}c@{}}Exploit \\ Mechanism\end{tabular}}} & \multicolumn{1}{c|}{\textbf{Payloads}} & \multicolumn{1}{c|}{\textbf{Vulnerability}} \\ \hline
\multicolumn{1}{|l|}{\begin{tabular}[c]{@{}l@{}}Tapo \\ Camera\end{tabular}} & \multicolumn{1}{l|}{\begin{tabular}[c]{@{}l@{}}443 (HTTPS), \\ 554 (RTSP), \\ 8800, 2020\end{tabular}} & \multicolumn{1}{l|}{Digest} & \multicolumn{1}{l|}{POST} & \multicolumn{1}{l|}{\begin{tabular}[c]{@{}l@{}}Forged HTTP \\ POST request \\ with Digest \\ headers\end{tabular}} & \multicolumn{1}{l|}{\begin{tabular}[c]{@{}l@{}}POST /stream with \\ crafted Digest fields\\ (username, nonce, etc.)\end{tabular}} & \multicolumn{1}{l|}{\begin{tabular}[c]{@{}l@{}}IoT Access\\ Credentials \\ Leakage\end{tabular}} \\ \hline
\multicolumn{1}{|l|}{\begin{tabular}[c]{@{}l@{}}Netatmo\\ Camera\end{tabular}} & \multicolumn{1}{l|}{80 (HTTP)} & \multicolumn{1}{l|}{------} & \multicolumn{1}{l|}{GET} & \multicolumn{1}{l|}{\begin{tabular}[c]{@{}l@{}}Forged HTTP\\ GET request \\ with URI path\end{tabular}} & \multicolumn{1}{l|}{\begin{tabular}[c]{@{}l@{}}GET /65fc8c8fcc8a5cde\\ 31795a6c94f606b3/live/\\ files/high/live0000000839.ts\end{tabular}} & \multicolumn{1}{l|}{\begin{tabular}[c]{@{}l@{}}Sneak IoT Live\\ Video Stream\end{tabular}} \\ \hline
\multicolumn{1}{|l|}{\begin{tabular}[c]{@{}l@{}}Kodak\\ Camera\end{tabular}} & \multicolumn{1}{l|}{80 (HTTP)} & \multicolumn{1}{l|}{Basic} & \multicolumn{1}{l|}{GET} & \multicolumn{1}{l|}{\begin{tabular}[c]{@{}l@{}}Forged HTTP\\ GET request \\ with Basic \\ and URL\end{tabular}} & \multicolumn{1}{l|}{\begin{tabular}[c]{@{}l@{}}1. GET/000008E048AF014EAB\\ VAXNGI /E048AF014EAB\_\\ 01\_20240105135201866.jpg?\\ 2. video/x-flv\end{tabular}} & \multicolumn{1}{l|}{\begin{tabular}[c]{@{}l@{}}1. Creep IoT live\\ images\\ 2. Sneak IoT Live\\ Video Stream\end{tabular}} \\ \hline
\multicolumn{1}{|l|}{\begin{tabular}[c]{@{}l@{}}Imou\\ Camera\end{tabular}} & \multicolumn{1}{l|}{\begin{tabular}[c]{@{}l@{}}80 (HTTP)\\ 554 (RTSP)\end{tabular}} & \multicolumn{1}{l|}{WSSE} & \multicolumn{1}{l|}{POST} & \multicolumn{1}{l|}{\begin{tabular}[c]{@{}l@{}}Forged HTTP\\ POST request\\ with WSSE profile\\ information\end{tabular}} & \multicolumn{1}{l|}{\begin{tabular}[c]{@{}l@{}}POST /device/7H04FF7PAZ2\\ D298/transfer-stream/real/0/0/3\end{tabular}} & \multicolumn{1}{l|}{\begin{tabular}[c]{@{}l@{}}Sneak IoT live \\ streams video\end{tabular}} \\ \hline
\multicolumn{1}{|l|}{\begin{tabular}[c]{@{}l@{}}Alarm\\ SpyClock\end{tabular}} & \multicolumn{1}{l|}{80 (HTTP)} & \multicolumn{1}{l|}{OSS} & \multicolumn{1}{l|}{PUT} & \multicolumn{1}{l|}{\begin{tabular}[c]{@{}l@{}}Forged HTTP\\ PUT request with \\ URI\end{tabular}} & \multicolumn{1}{l|}{\begin{tabular}[c]{@{}l@{}}1. PUT /AYS-210677-MHCEX/\\ photos/20240118/2024-01-18-15\\ -02-3-0.jpg\\ 2. PUT/AYS-210677-MHCEX/\\ dates/20240118\end{tabular}} & \multicolumn{1}{l|}{\begin{tabular}[c]{@{}l@{}}Creep IoT live\\ images\end{tabular}} \\ \hline
\multicolumn{1}{|l|}{\begin{tabular}[c]{@{}l@{}}Airveda\\ Air\\ Purifier\end{tabular}} & \multicolumn{1}{l|}{\begin{tabular}[c]{@{}l@{}}80 (HTTP)\\ TCP\end{tabular}} & \multicolumn{1}{l|}{-----} & \multicolumn{1}{l|}{GET} & \multicolumn{1}{l|}{\begin{tabular}[c]{@{}l@{}}Forged HTTP\\ GET request\end{tabular}} & \multicolumn{1}{l|}{\begin{tabular}[c]{@{}l@{}}GET /core/getDeviceDetails\\ /?deviceUid=1212210161\end{tabular}} & \multicolumn{1}{l|}{\begin{tabular}[c]{@{}l@{}}IoT Access\\ Credentials\\ Leakage\end{tabular}} \\ \hline
 &  &  &  &  &  & 
\end{tabular}}
\end{table*}

To investigate the generalizability of IoTFuzzSentry, we extend our analysis of the traffic of six additional IoT devices, including four IoT cameras, namely Tapo, Netatmo, Kodak, and Imou, a spy clock, namely Alarm Spy Clock, and an air quality monitor, namely Airveda Air Purifier.

All these IoT devices use the HTTP/RTSP protocol with TCP at the transport layer for communication with their respective cloud servers. The payloads in the HTTP packets contain the exact URIs, for example, to denote the resources in these IoT devices. More importantly, the user authentication tags in these payloads are \textit{Basic} or \textit{Digest} or \textit{WSSE} or \textit{OSS}, any of these contains encoded user credentials that can be easily decoded. A summary of the traffic analysis of these six additional devices, together with the respective payloads that can be fuzzed, is shown in Table \ref{tab:gen_6iotdevices}. 
Developing a separate module for each of these devices for the integration into IoTFuzzSentry is considered a part of our immediate future work.

\subsection{Limitation of IoTFuzzSentry}
In its current capabilities, we have considered only plaintext payloads, e.g., HTTP or RTSP, for applying the fuzzing.  One of the IoT cameras, namely Tapo Camera, uses the HTTPS protocol for some of its communications along with the RTSP protocol. 
IoTFuzzSentry is currently not capable of exploiting encrypted payloads. 
While encryption prevents direct inspection of the payloads, certain existing research works, such as those in \cite{Fiteruau2020dtlsfuzzing}, \cite{Zhang2024protocolfuzzing}, and \cite{ang2025quicfuzzeffectivegreyboxfuzzer}, demonstrate that fuzzing at the application layer, before encryption, may still be possible. 
In particular, some control commands may still be transferred by using easy-to-break encryption protocols. 
We consider any such exploration of encrypted payloads as part of our future works. It is worth noting that a number of existing IoT fuzzers, like BooFuzz \cite{boofuzz}, Peach \cite{PeachFuzzer}, Randamsa \cite{Radamsa}, and SPIKE \cite{SPIKE}, are limited to plaintext payloads and the current limitations are in line with these existing fuzzers.

\subsection{Comparing with Existing Tools}
\label{sec:compare-existing-works}
To investigate if the existing fuzzers can be suitable to discover the reported vulnerabilities in their current capabilities, we have considered four state-of-the-art fuzzers, namely Fuzzoswki, AFLNet, Snipuzz, and Doona; results are summarized in Table \ref{tab:comp_stateofarttools}. Because the main is to detect crash, Fuzzoswki does not consider the response produced by IUT. 
With a similar aim as that in Fuzzoswki, ALFNet requires a sequence of packets as seed, and it turns out that IUT may accept one packet but discard others in a given sequence.
The existing payloads in Doona cause bad requests as a response from IUT. 
Snipuzz actually requires certain API in a firmware for which the seed inputs are structured, where it basically looks for crash. Hence, we claim that the vulnerabilities reported in this paper have not been seen before or the existing tools are not suitable to discover them in their current capabilities.

\section{Conclusion}
\label{sec:conclusion}
In this paper, we address the problem of discovering security vulnerabilities in operational IoT devices using protocol-guided fuzzing. In particular, we have designed and developed a mutation-based protocol fuzzer called \textit{IoTFuzzSentry}. It 
automatically extracts protocol-specific payloads from past network traces of an IoT device under test (IUT). This is then used to generate mutated payloads based on the protocol and IUT. Currently, the tool is highly customizable and has been applied to three IoT application protocols, including two well-known protocols (HTTP and RTSP) and one proprietary protocol (TP-Link SmartHome/JSON). Our experiments revealed 
seven new vulnerabilities, showing the effectiveness of \textit{IoTFuzzSentry}. 
Our approach is extensible to other commercially available IoT devices with minimal modification.
We hope that \textit{IoTFuzzSentry} provides a foundation for IoT fuzzing in finding security vulnerabilities beyond denial-of-service.

\section*{Ethical Consideration}
The authors state that no sensitive information about any users or vendors was collected or stored, and that no harm was caused to real-world systems or services during testing process.

\bibliographystyle{IEEEtran}
\bibliography{references}

\end{document}